\documentclass[10 pt, journal]{IEEEtran}

\makeatletter
\def\ps@headings{%
	\def\@oddhead{\mbox{}\scriptsize\rightmark \hfil \thepage}%
	
	\def\@evenhead{\scriptsize\thepage \hfil \leftmark\mbox{}}%
	
	\def\@oddfoot{}%
	
	\def\@evenfoot{}}
	
\makeatother
\usepackage[hidelinks]{hyperref}
\usepackage{cite}
\usepackage[table]{xcolor}

\usepackage{float}  
\usepackage{epsfig,bm,amsmath,amssymb,graphicx,setspace}
\usepackage[numbers,sort&compress]{natbib}
\usepackage{color,placeins}
\usepackage{multirow}
\usepackage{algorithm}
\usepackage{algpseudocode}
\usepackage{colortbl}
\usepackage{color}
\usepackage{dblfloatfix}
\usepackage{turnstile}
\usepackage{multicol}
\usepackage{array}

\usepackage{enumerate}
\usepackage{turnstile}
\usepackage{subcaption}
\usepackage{arydshln,paralist}
\usepackage{soul,tabularx,booktabs}
\usepackage{csquotes}
\usepackage{multirow}
\usepackage{adjustbox}
\usepackage{float}

\usepackage{xcolor}
\usepackage{verbatim}
\usepackage[colorinlistoftodos]{todonotes} 
\usepackage{rotating}
\definecolor{usethiscolorhere}{rgb}{0.86666,0.78431,0.78431}
\usepackage{tikz}
\usetikzlibrary{mindmap}
\usetikzlibrary{calc,positioning}
\usepackage{lipsum,adjustbox}
\usetikzlibrary{decorations.pathmorphing}

\usepackage{pifont}

\makeatother

\usepackage{amsmath}
\DeclareMathOperator{\clamp}{clamp}
\DeclareMathOperator{\round}{round}
\begin{document}

\title{FAST-IDS: A Fast Two-Stage Intrusion Detection System with Hybrid Compression for Real-Time Threat Detection in Connected and Autonomous Vehicles}

\author{Devika S, Vishnu Hari, Pratik Narang,~\IEEEmembership{Senior Member, ~IEEE}, Tejasvi Alladi, ~\IEEEmembership{Senior Member, ~IEEE}, and Vinay Chamola, ~\IEEEmembership{Senior Member, ~IEEE}

\thanks{Devika S, Vishnu Hari, Pratik Narang and Tejasvi Alladi are with the Department of Computer Science and Information Systems, BITS Pilani, Pilani Campus, 333031, India.  (e-mail: p20210024@pilani.bits-pilani.ac.in; f20220094@pilani.bits-pilani.ac.in; pratik.narang@pilani.bits-pilani.ac.in; tejasvi.alladi@pilani.bits-pilani.ac.in).}

\thanks{Vinay Chamola is with the Department of Electrical and Electronics Engineering \& APPCAIR, BITS Pilani, Pilani Campus, 333031, India.  (e-mail: vinay.chamola@pilani.bits-pilani.ac.in).}
}
\maketitle{}

\begin{abstract}

 In the realm of Intelligent Transportation Systems (ITS) and Connected and Autonomous Vehicles (CAVs), the ever-growing connectivity among vehicles, roadside infrastructure, and the Internet opens the door to a wide array of cybersecurity threats. Intrusion Detection Systems (IDSs) have thus been essential for ensuring safety and efficiency in ITS.  While prior studies highlight the importance of detecting novel attacks, adopting multi-stage IDS architectures, and maintaining computational efficiency, only a few frameworks address all these aspects collectively, and many overlook deployment in resource-constrained environments. To address these gaps, we introduce a novel two-stage IDS optimized through hybrid model compression, integrating structural pruning and static quantization, achieving a 77.2\% reduction in model size while maintaining computational efficiency. Validated on the VeReMi Extension dataset, the lightweight model supports real-time deployment on RTX A6000, Colab CPU, and Jetson Nano, achieving per-vehicle attack detection within 0.195 seconds, achieving approximately 50.05\% reduction in inference time. The first stage employs a coarse-grained 4-layer Convolutional Neural Network (CNN) integrated with a Bidirectional Generative Adversarial Network (BiGAN) to identify normal versus anomalous vehicular data, achieving a recall rate of 92.79\%. The second stage employs a fine-grained hybrid CNN–Long Short-Term Memory (CNN–LSTM) classifier, featuring a 2-layer CNN and a 1-layer LSTM, to discriminate among 19 known attack types with an accuracy rate of 97.822\%. By leveraging reconstruction error metrics and combining supervised with unsupervised techniques, the proposed framework significantly enhances the detection of zero-day attacks. Experimental evaluations show that the proposed IDS surpasses existing methods, offering a robust and scalable cybersecurity solution for next-generation intelligent transportation.

\end{abstract}


\section{Introduction}

Connected and Autonomous Vehicles (CAVs) are integrated networks that facilitate communication with other vehicles within a specified range and exhibit varying levels of autonomy. The development of CAVs has been motivated by enhanced road safety, advancements in communication technology, economic benefits and user requirements, thereby achieving the objectives of Intelligent Transport Sytems (ITSs) \cite{kumari2023safety,alladi2021artificial}. The full functionality of CAVs relies on continuous connectivity within a specified range, using the IEEE 802.11 standard for the allocated 5.9 GHz frequency band \cite{festag2014cooperative}. The IEEE 802.11 standard, which resembles Wi-Fi technology, enables vehicles to connect with other vehicles (Vehicle-to-Vehicle), roadside infrastructure (Vehicle-to-Infrastructure), cloud servers (Vehicle-to-Cloud), pedestrians (Vehicle-to-Pedestrian) and essentially everything else (Vehicle-to-Everything) \cite{shladover2018connected}.

According to the World Health Organization (2019), traffic accidents accounted for 93\% of all fatalities, resulting in 1.3 million deaths. While CAVs can significantly reduce accidents caused by human error, their increasing connectivity through Basic Safety Messages (BSMs) exposes them to privacy risks and cyber threats. Consequently, robust solutions, including Blockchain technology \cite{kumar2024survey}, cryptography-based encryption \cite{xiao2024calra}, and authentication schemes \cite{ismail2024designing,akram2023fog}, have been developed to protect CAVs.

However, encryption techniques alone proved inefficient, introducing an Intrusion Detection System (IDS). IDS proved crucial to complement encryption-based security \cite{alladi2023deep} by providing network-wide monitoring and protecting individual communication links in vehicular networks. The dynamic and heterogeneous nature of cyber threats necessitates a robust IDS framework \cite{alladi2023deep}, with deep learning models being well-suited due to the vast data exchange among vehicles. IDS have evolved significantly, progressing from plausibility checks \cite{kamel2020veremi} to advanced deep learning models \cite{devika2024vadgan,sreelekshmi2025deep}.

Recent studies implemented on hybrid deep learning models and generative AI models have effectively addressed challenges in IDS. For instance, \cite{sajid2024enhancing} combined machine learning and deep learning models to improve intrusion detection on benchmarked datasets, though several of the datasets used were outdated. Whereas, \cite{sreelekshmi2025deep} proposed an optimized CNN architecture targeting physical-layer message features. 
Additionally, multi-stage intrusion detection \cite{alladi2023deep,lee2008multi,zhou2021detecting,chen2023provably} enhances performance compared to single-stage classifiers and also addresses computational latency, which is vital for efficient operation in resource-constrained vehicular environments \cite{kim2022vehicular, lee2023malicious}. Specifically, \cite{althunayyan2024robust} demonstrated multi-stage detection with an ANN for Stage 1 and LSTM being utilized in Stage 2, demonstrating effectiveness against previously unseen attacks. 
Generative AI, particularly Generative Adversarial Networks (GANs), was proven effective in detecting anomalies through its reconstructive properties \cite{devika2024vadgan, xie2021threat}, enabling the identification of unseen attacks \cite{zhao2022can}. 
 However, despite these advantages, they fall short in comprehensively addressing zero-day attack detection, a lightweight model that can be deployed in a resource-constrained environment and achieving fast real-time response.  
Motivated by these challenges, we developed a two-stage IDS utilizing Bidirectional Generative Adversarial Network (BiGAN) \cite{donahue2016adversarial} with Wasserstein loss and gradient penalty in Stage 1 and hybrid CNN-LSTM in Stage 2. Our two-stage model integrates model compression to minimize computational latency and thereby effective for deployment on resource-constrained platforms while ensuring high detection accuracy and fast inference.

The contributions of this study are:
\begin{enumerate}[i.]
\item
Two-Stage Intrusion Detection Framework: We utilized the capabilities of BiGAN for coarse-grained anomaly detection (Stage 1) and CNN-LSTM for fine-grained classification (Stage 2), resulting in a robust hybrid model that significantly enhances the security in CAVs.
\item
Hybrid Model Compression: We proposed a model compression technique combining structural pruning with static quantization, achieving a 50.05\% reduction in inference time and a 77.2\% reduction in model size, with approximately 5\% performance loss. The inference performance was evaluated on RTX A6000 GPU, Colab CPU, and Jetson Nano, with a per-vehicle detection time of 0.195 seconds in real time.
\item
Detection of Previously Unseen Attacks: Stage 1 applied unsupervised metrics (MSE, Mahalanobis distance) for anomaly detection, while Stage 2 used supervised learning to refine detection and identify unseen attacks. This hybrid supervised-unsupervised approach improved performance, achieving 85.05\% accuracy on unseen threats while maintaining strong detection of known attacks.
\item
Performance Metrics Evaluation: We achieved a recall rate of 92.785\% in the first stage for classifying data as normal or anomalous, and accuracy rate of 97.822\% in the second stage for classifying anomalous data into 19 known attack types, surpassing existing state-of-the-art (SOTA) methods.

\end{enumerate}

The rest of this paper is organized as follows: Section \ref{sec:literature} surveys the related works. Section \ref{sec:background} provides a preliminary background, including the CAV scenario, the dataset used, and the taxonomy of attacks within the scope of this study.
\textcolor{black}{Section \ref{sec:proposed} outlines the proposed intrusion detection architecture, detailing Stages 1 and 2, followed by the hybrid data compression implementation. Section \ref{sec:simulation} describes the experimental setup, while Section \ref{sec:results} presents the results and analysis. Finally, Section \ref{sec:conc} concludes the study.}

\section{Related Work}
\label{sec:literature}

\begin{table*}[t]
\centering
\caption{Comparative Study}
\label{tab: literature survey}
\resizebox{1.7\columnwidth}{!}{%
\begin{tabular}{ccccccc}
    \toprule
    \textbf{S. No.} & \textbf{Ref.} & \textbf{Dataset - Attacks} & \textbf{Models Used} &\multirow{2}{15mm}{\textbf{Multistage Detection} }& \multirow{2}{15mm}{\textbf{Unseen Attacks}} &
    \multirow{2}{35mm}{\textbf{Is Computational Efficiency talked about ?}} \\\\\\
    \midrule
    1 & \cite{alladi2021artificial} & VeReMi Extension-19 & Hybrid cnn-lstm & X & X & \checkmark \\
    2 & \cite{amanullah2022burst} & BurST-ADMA-7 &Supervised ML & X & X & X \\
    3 & \cite{fan2024auto} & VeReMi - 8
 & CNN + Self-attention & X & \checkmark & \checkmark \\  
     4 & \cite{sharma2020machine} & VeReMi - 5
 & Supervised ML, Ensemble & X & X & X \\  
      5 & \cite{sharma2019attacks} & VeReMi - 5
 & DL & X & \checkmark & X \\  
 6 & \cite{alladi2023deep} & VeReMi Extension-17
 & Hybrid cnn-lstm & \checkmark & X & \checkmark \\ 
 7 & \cite{devika2024vadgan} & VereMiExtension-19 & GAN & X & \checkmark& \checkmark\\
  8 & \cite{xie2021threat} & CAN-10 & GAN & X & \checkmark& X \\
  9 & \cite{ullah2022hdl} & Car Hack2020 - 5& LSTM, GRU & X & X& X\\
  10 & \cite{chougule2022multibranch} & VeReMi Extension-19 & CNN + Reconstruction & X & \checkmark& \checkmark\\
\bottomrule
\end{tabular}
}
\label{tab:fim}
\end{table*}

This section presents a comprehensive review of existing intrusion detection research in CAVs. Table  \ref{tab: literature survey} provides a detailed comparison, highlighting datasets used, attack coverage, multistage detection capability, zero-day attack identification, and computational latency considerations.

\subsubsection{Intrusion Detection Systems in CAVs}

The deployment of Hybrid CNN-LSTM models on Multi-access Edge Computing (MEC) servers for attack classification was explored in \cite{alladi2021artificial}, utilizing LSTMs for sequence-based classification and CNNs for sequence-image classification, enabling real-time processing. The study in \cite{amanullah2022burst} introduced the Burwood SUMO Traffic (BurST) dataset, simulating Australian traffic conditions and employing supervised learning algorithms such as KNN, SVM, Naive Bayes, and Random Forest for attack classification. In \cite{sharma2020machine}, plausibility checks, including Location and Movement Plausibility, were integrated into supervised learning models for attack detection in CAVs. Authors in \cite{chougule2022multibranch} segmented vehicular data \cite{kamel2020veremi} into frequency (timestamp), identity (pseudo-identity), and motion, processing them through three CNN-based reconstruction models with branching and reconstruction error thresholds.
\subsubsection{Computational Efficiency in Resource Constrained Environments}
The study in \cite{kim2022vehicular} addressed computational latency by compressing Controller Area Network (CAN) messages through XOR-based encoding, transmitting only the differences between consecutive frames while utilizing the dedicated Compression Area Map (CAP) to preserve data integrity. Similarly, \cite{lee2023malicious} adopted a bitmap-based compression method for malicious traffic, incorporating a predefined threshold and a relearning step that reduced memory consumption and training time. While \cite{chougule2022multibranch} proposed a CNN-based reconstruction approach for IDS, emphasizing deployment in resource-constrained environments such as Jetson Nano and Google Colab, with applicability to edge and cloud platforms. Another notable improvement is presented in \cite{tutsoy2024novel}, where the authors have performed feature selection, handling of missing data, and dimensionality reduction using a deep machine learning algorithm. Although this work is from the medical domain, it is highly relevant to our work and provides a valuable direction for the future. Though effective in computing, their optimizations are mainly restricted to the data level or struggle to achieve robust performance in real-world implementations.

\subsubsection{GANs for Intrusion Detection}
Authors in \cite{devika2024vadgan} propose an unsupervised GAN framework for attack detection in CAVs, leveraging reconstruction properties to detect previously unseen attacks while focusing on motion-related attributes from the VeReMi Extension dataset. The authors of \cite{xie2021threat} present a GAN-based intrusion detection method for automotive CAN networks, enhancing attack detection by integrating the CAN communication matrix to identify threats like DoS, injection, masquerade, and data tampering. The work in \cite{zhao2022can} introduces a multi-stage GAN-based IDS for CAN networks, incorporating an Auxiliary Classifier GAN (ACGAN) and Out-of-Distribution (OOD) detection to improve identification of both known and unknown attacks. While these reconstruction-based mechanisms claim the detection of previously unseen attacks, their detection capability remains uncertain. 

\subsubsection{Detection of Unseen Attacks}
The research in \cite{fan2024auto} utilizes a CNN and a self-attention mechanism for feature extraction. Although the authors claim to detect new attacks over time using their update mechanism, they do not explicitly show any results. Similarly, the authors in \cite{sharma2019attacks} have demonstrated that models such as Logistic Regression and LSTM failed to detect adversarial perturbations and did not devise any IDS capable of detecting them. Implementing GANs \cite{devika2024vadgan,xie2021threat,zhao2022can} in intrusion detection enhances the detection of previously unseen attacks by training the generator to create adversarial examples and the discriminator to recognize new attack patterns. While they have achieved good performance in detecting known attacks, they have not shown the detection of previously unseen attacks.
\subsubsection{Multistage Intrusion Detection}
The work in \cite{alladi2023deep} implements three Deep Learning Classifier Engines (DLCE-1, DLCE-2, DLCE-3) in single-stage and multi-stage configurations on edge servers, enhancing security and reducing computational latency using Hybrid CNN-LSTM models. A multi-stage intrusion detection system in \cite{lee2008multi} utilizes a Hidden Markov Model (HMM) to identify intrusions through characteristic signals while distributing workload to minimize overhead. The LSTM-based autoencoder proposed in \cite{zhou2021detecting} captures long-term dependencies for multi-stage attack detection across four public datasets, with a multi-encoder approach improving efficiency for real-time intrusion detection. However, the VeReMi extension dataset remains underutilized in multi-stage IDS, and effective integration of factors such as unseen attack detection and lightweight model deployment in resource-constrained environments is still lacking.

\captionsetup[figure]{skip=0.5pt}
\begin{figure}
    \centering
    \includegraphics[width=0.45\textwidth]{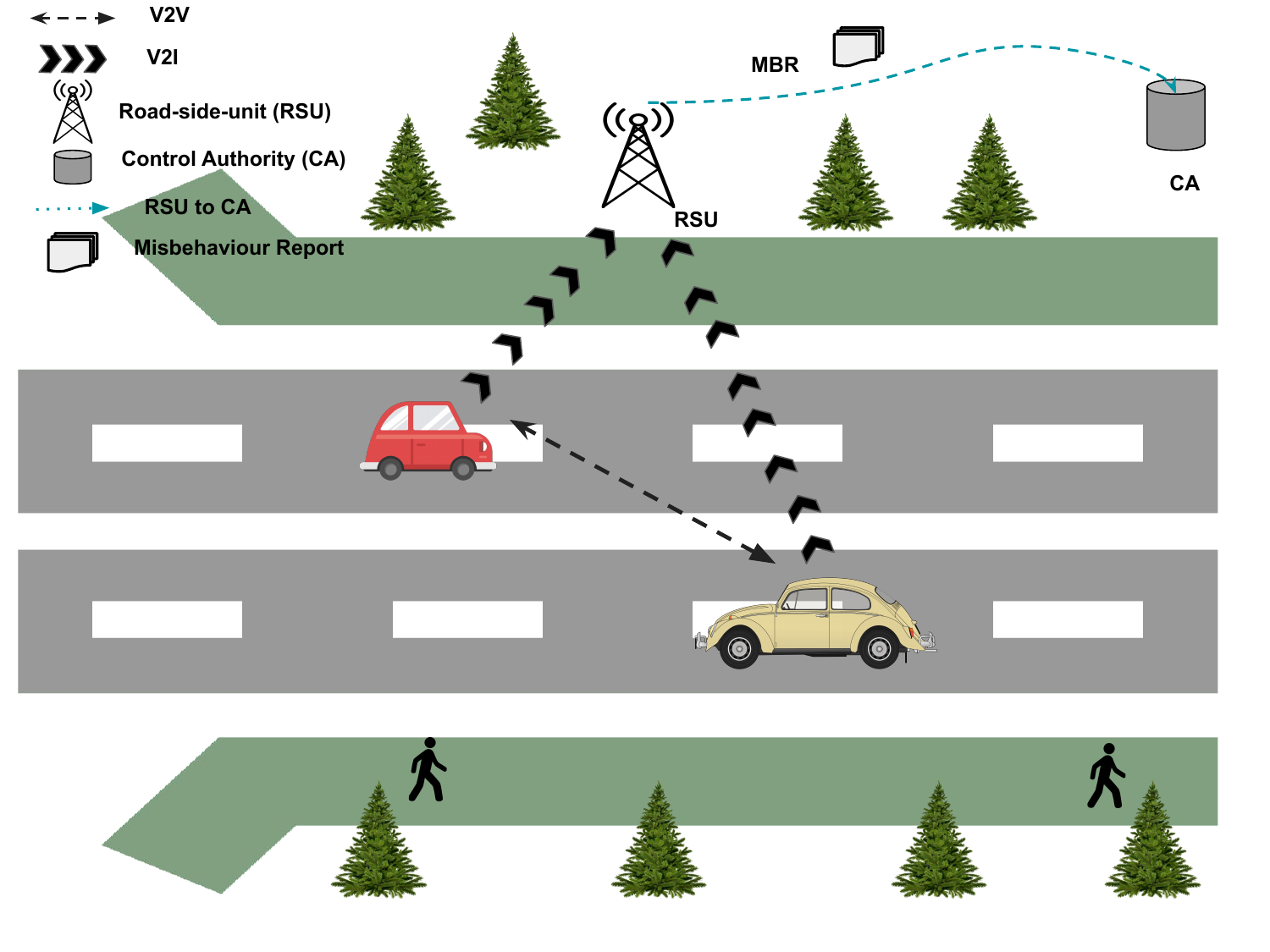}
    \caption{A typical CAV scenario.}
    \label{fig:IoV}
\end{figure}

\textcolor{black}{Our comprehensive literature review underscored the need for an IDS that integrates model compression to mitigate computational latency, generative AI to exploit reconstructive properties for detecting previously unseen attacks, and a multistage architecture to capture long-term dependencies in time-series data. These insights led to developing our two-stage intrusion detection model, providing a fast and efficient real-time solution. }

\begin{figure*}
    \centering    \includegraphics[width=0.9\textwidth]{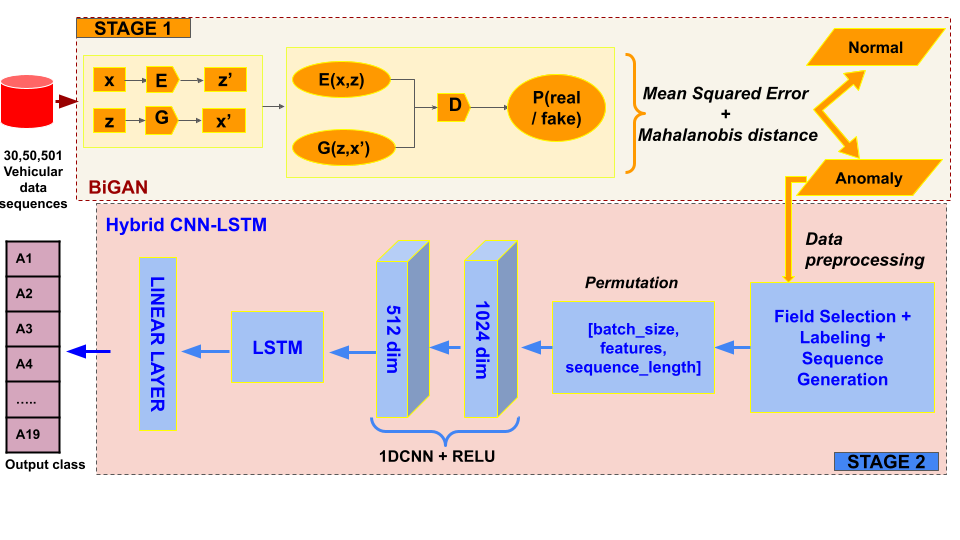}
    \caption{Proposed FAST-IDS architecture.}
    \label{fig:arch}
\end{figure*}

\section{Preliminary Background}
\label{sec:background}
\subsubsection{CAV Scenario}
Fig.~\ref{fig:IoV} illustrates a typical CAV scenario where vehicles communicate via Vehicle-to-Vehicle (V2V) and Vehicle-to-Infrastructure (V2I) networks.  Each vehicle processes GPS and sensor data locally through an On-Board Unit (OBU).  The vehicles must register with a Control Authority (CA) for authentication within its operational area \cite{alladi2023deep}.  Roadside Units (RSUs) manage message transmission and intrusion detection, analyzing received data to detect potential attacks.  A Misbehaviour Report (MBR) is generated and sent to the CA if a message is flagged as anomalous. Based on the received MBRs, the CA can take appropriate actions, such as revoking the vehicle’s certificate, to mitigate security threats if necessary.

\subsubsection{Dataset}

The VeReMi Extension dataset \cite{kamel2020veremi}, an enhancement of the original VeReMi dataset \cite{van2018veremi}, is used in our IDS evaluation. The dataset is generated using the Framework for Misbehaviour Detection (F2MD), an extension of VEINS, integrating OMNeT++ for network simulation and SUMO \cite{sommer2019veins} for traffic simulation, and is validated using real-vehicular data from Luxembourg. Realistic sensor error models for position, speed, acceleration, and heading ensure accurate simulation based on real GPS and vehicular data. The dataset includes sender vehicle ID, timestamp, pseudo-identity, message ID, and vehicle coordinates. Plausibility and consistency checks further validate its reliability for misbehaviour detection \cite{kamel2020veremi}.

\subsubsection{Taxonomy of Attacks}

The dataset includes 19 attacks that span different traffic densities (high/low traffic density) and times of day (24 hours).
The attack names, class labels assigned to each attack and their descriptions are detailed in \cite{devika2024vadgan}.

\section{Proposed Intrusion Detection Framework}
\label{sec:proposed}
The proposed FAST-IDS architecture is illustrated in Fig. \ref{fig:arch}. The architecture consists of two stages. The first stage performs coarse-grained classification, distinguishing between normal and anomalous data, and the second stage conducts fine-grained classification, identifying the input sequence as one of 19 known attack types. We have further compressed this model to be deployed in resource-constrained environments.

\subsection{Stage 1}
\subsubsection{Data Preprocessing}

We have utilized the  24-hour benchmarked VeReMi Extension dataset \cite{kamel2020veremi}, selecting the motion-related information in line with prior studies \cite{kushardianto2024vehicular,hsu2021deep,devika2024vadgan}. Each instance or sequence is denoted as $Seq_{i}$, where the $i^{th}$ sequence includes features such as $(p_{x_i}, p_{y_i}, s_{x_i}, s_{y_i}, a_{x_i}, a_{y_i}, h_{x_i}, h_{y_i})$, corresponding to position, speed, acceleration, and heading in the x and y direction, respectively.

\subsubsection{Architectures Employed}

 Each input sequence \(Seq_{i}\) is fed into the encoder $E$, a 5-layer CNN, which produces the latent code $E(x)$. This latent code is then fed into the generator $G$, a 4-layer CNN, which reconstructs the input sequence as \(Seq'_{i}\). The discriminator $D$, also implemented as a 4-layer CNN, receives joint pairs from the encoder $(x, E(x))$ and the generator $(G(E(x), E(x))$, and outputs a probability indicating whether the data originates from the encoder (real) or the generator (fake). The model is trained using the Wasserstein loss with a gradient penalty to enforce the Lipschitz constraint on discriminator weights.

\begin{equation}
\label{eq:obj}
\min_{G,E} \; \max_D \; V(D, G, E),
\end{equation}

where the value function $V(D,G,E)$ is given by

\begin{equation}
\label{eq:2}
\begin{aligned}
V(D,G,E) = \; & 
\mathbb{E}_{x \sim p_\text{data}} \big[ D(x, E(x)) \big] 
- \mathbb{E}_{x \sim p_\text{data}} \big[ D(G(E(x)), E(x)) \big] \\
& + \lambda_{gp} \, \mathbb{E}_{\hat{x} \sim p_{\hat{x}}} 
\Big[ \big( \| \nabla_{\hat{x}} D({G(E(x))}, E(x)) \|_2 - 1 \big)^2 \Big].
\end{aligned}
\end{equation}

Here, $\mathbb{E}_{x \sim p_{\text{data}}}$ refers to the expectation over data and $\lambda_\text{gp}$ is the gradient penalty term.

\subsubsection{Sequence Reconstruction and Reconstruction Error Calculation}
\label{sec:re}

The \(Seq'_{i}\) reconstructed by the generator corresponds to \((p'_{x_i}, p'_{y_i}, s'_{x_i}, s'_{y_i}, a'_{x_i}, a'_{y_i}, h'_{x_i}, h'_{y_i})\), representing the reconstructed values of position, speed, acceleration, and heading in the x and y directions, respectively. The Reconstruction Error (RE) was measured using weighted sum of Mean Squared Error (MSE) and Mahalanobis Distance (\(D_M\)) \cite{xu2018joint}. 

\begin{equation}
    MSE = \frac{1}{n}\sum_{i=1}^{n}{((Seq_{i})-(Seq'_{i}))^2}
\end{equation}

\begin{equation}
D_M = \sqrt{(\mathbf{ \text{Seq}'_i} - \mu)^T \Sigma^{-1} (\mathbf{ \text{Seq}'_i} - \mu)}
\end{equation}
Here, \(\mu\) is the mean vector of the input data, \(\Sigma^{-1}\) is the inverse of the covariance matrix of the input data.

The combined score (RE) is calculated as follows:
\begin{equation}
\text{Combined Score} = \alpha \cdot D_M + (1 - \alpha) \cdot \text{MSE}
\end{equation}

Here, $\alpha$ = 0.5 is a weighting factor that balances the contributions of the Mahalanobis distance and the MSE.
The Interquartile Range (IQR) was employed to determine the range of values within which the RE of the ground truth data lies.

\begin{equation}
\text{IQR} = Q3 - Q1
\end{equation}

\begin{equation}
\text{Upper Limit (UL)} = Q3 + 1.5 \times \text{IQR}
\end{equation}

\begin{equation}
\text{Lower Limit (LL)} = Q1 - 1.5 \times \text{IQR}
\end{equation}
Q1 corresponds to the 25th percentile, whereas Q3 corresponds to the 75th percentile. We have utilized different quartiles to distinguish between anomalous and normal data.
\begin{align}
    \text{For ANOMALY} &= 
    \begin{cases}
        1, & \text{for } RE <  Q1 \text{or} RE > Q3 \\
        0, & \text{otherwise}
    \end{cases} \\
    \text{For NORMAL} &= 
    \begin{cases}
        1, & \text{for } \text{LL} \leq RE \leq \text{UL} \\
        0, & \text{otherwise}
    \end{cases}
\end{align}

Here, LL and UL are linearly adjusted forms of Q1 and Q3, respectively. 
\subsection{Stage 2}
\subsubsection{Data Preprocessing}
The data sequences \cite{kamel2020veremi} flagged as anomalous from Stage 1 are fed as input to Stage 2 \cite{chougule2022multibranch}. To prepare the input data, we create fixed-length window sequences of size $w$, inspired by~\cite{alladi2021artificial}. This approach produces sequences containing $w$ consecutive samples, ensuring temporal order. Each sample \(Seq_\text{iA}\) is of the form \((p_{x_i}, p_{y_i}, s_{x_i}, s_{y_i}, a_{x_i}, a_{y_i}, h_{x_i}, h_{y_i})\), representing the features same as in Stage 1. As illustrated in Fig. \ref{fig:arch}, match the input expectations of the Hybrid CNN-LSTM model, the data is rearranged into the format ($batch\_size$, $features$, $sequence\_length$).

\subsubsection{Architectures Employed and Multiclass Classification}

The input data \(Seq_\text{iA}\) is fed into a 2-layer CNN with ReLU activation to extract features. The generated latents are then passed into an LSTM,  which captures temporal dependency. A dense head, represented by a single linear layer with softmax activation, maps these latents to class prediction probabilities. We have also experimented with the model using Bidirectional LSTM variant and various loss functions, including Cross-Entropy loss (CE), Label Smoothing Loss (SL), and Focal Loss (FL). The results are detailed in Section \ref{sec:results}. 

\begin{equation}
CE = - \frac{1}{N} \sum_{n=1}^{N} \sum_{i=1}^{C} y_i^n \log(p_i^n)
\end{equation}
\label{ce}

\begin{equation}
\label{eq:sl}
SL = - \frac{1}{N} \sum_{n=1}^{N} \sum_{i=1}^{C} \left( (1 - \alpha) y_i^n + \frac{\alpha}{C} \right) \log(p_i^n)
\end{equation}

\begin{equation}
\label{eq:fl}
FL = - \frac{1}{N} \sum_{n=1}^{N} \sum_{i=1}^{C} (1 - p_i^n)^\gamma y_i^n \log(p_i^n)
\end{equation}

Here, $N$ is the number of data points in the batch and $C$ is the number of classes. $y_i^n$  and $p_i^n$ corresponds to the true and the predicted probability of the $n$-th data point for class $i$ respectively. In Eq. \ref{eq:sl}, $\alpha$ is the smoothing parameter, which spreads probability across all classes and in Eq. \ref{eq:fl} $\gamma$ is the focus parameter that helps reduce the weights on the parameters. 

\subsection{Hybrid Model Compression}
\textcolor{black}{We implemented a two-stage model compression technique to enhance model efficiency. First, structured pruning was applied to convolutional layer filters using L1 importance with a pruning ratio, followed by fine-tuning for $E$ epochs. Next, static quantization was performed on weights and activations using a min-max observer, with calibration conducted on a subset of the training dataset. We have utilized asymmetric quantization for activation and symmetric quantization for weights as suggested in prior studies \cite{nagel2021white}.} The formula for structured pruning is as follows:
\begin{equation}
  I_w = \sum |w^{pre}|   
\end{equation}

Here $I_w$ is the importance score of filter, $W^\text{pre}$ is the weights between the layers before compression. $| W^\text{pre} |$ gives the absolute value (L1 norm), after which we sort the filters in ascending order, discarding the least important weights based on pruning ratio $p$. After which, the model is fine-tuned for $E$ epochs and then fed to the quantization module.

The static quantization formula for weights $w$ and activations $x$ are as follows:

\begin{equation}
\begin{aligned}
 w^{q} &= \clamp\!\left(
            \round\!\left(\frac{w^{\mathrm{ft}}}{s}\right),
            -2^{n-1},\, 2^{n-1}-1
          \right) \\
  x^{q} &= \clamp\!\left(
            \round\!\left(\frac{x}{s}\right) + z,
            0,\, 2^{n} - 1
          \right)
\end{aligned}
\end{equation}

where, Scale Factor ($s$) and Zero-Point ($z$) are defined as follows:
\begin{equation}
    s = \frac{w_{\max} - w_{\min}}{2^{n} - 1}
\end{equation}

\begin{equation}
    z = \text{round} \left(\frac{-w_{\min}}{s} \right)
\end{equation}

Here $s$ and $z$ are used to calculate the quantized weights ($w^q$) and activations ($x^q$) corresponding to $n$ bit width. The value of $s$ and $z$ depends on the weight tensors [$w_\text{min}$,$w_\text{max}$] and on the activation tensors [$x_\text{min}$,$x_\text{max}$], which are calculated during calibration. During inference, we dequantize the values utilizing the formula given below:

\begin{equation}
\label{eq:dequant}
{V_{\text{dequantized}} = s \times (V_\text{quantized} - z)}
\end{equation}

The Eq. \ref{eq:dequant} is applicable for both $w^q$ and $x^q$, where $V_\text{quantized}$ = $w^q$ or $x^q$. For symmetric quantization corresponding to weight tensors, z will be zero. The algorithm for the proposed FAST-IDS is outlined as follows in Algorithm~\ref{alg:FAST-IDS} and Algorithm \ref{alg:compression}.

\begin{algorithm}
\caption{FAST-IDS Algorithm}
\small
\label{alg:FAST-IDS}
\begin{algorithmic}[1]
\State \textbf{Stage 1 of FAST-IDS ALG}
\State \textbf{Inputs:} $x$ = input data, $z$ = noise vector
\State // $E$ = Encoder, $G$ = Generator, $D$ = Discriminator
\State // $\alpha$ = weighting factor
\State \textbf{Output:} $\hat{y}$ = Final assigned class label.

\For{\textbf{each training iteration}}
    \State Sample real data $x \sim p_{\text{data}}(x)$
    \State Encode real data: $\hat{z} \gets E(x)$
    \State Generate fake sample: $\hat{x} \gets G(\hat{z})$
    \State Train $D$ on real pair $(x, \hat{z})$ and fake pair $(\hat{x}, \hat{z})$
    \State Update $D$ by \textbf{maximizing} discriminator objective function (Eq.~\ref{eq:2})
    \State Update $E$ and $G$ by \textbf{minimizing} generator and encoder objective function (Eq.~\ref{eq:2})
\EndFor

\Statex
\State Compute Reconstruction Error (RE) as:
\State \hspace{1em} $RE = \alpha \cdot D_m + (1 - \alpha) \cdot \mathrm{MSE}$, where:
\State \hspace{1em} $\mathrm{MSE}(x, \hat{x}) = \frac{1}{n} \sum_{i=1}^{n} (x_i - \hat{x}_i)^2$
\State \hspace{1em} $D_m(x') = (x' - \mu)^T \Sigma^{-1} (x' - \mu)$

\Statex
\State Compute percentiles: $Q1 = 25^\text{th}$ percentile, $Q3 = 75^\text{th}$ percentile
\State Compute IQR: $\mathrm{IQR} = Q3 - Q1$
\State Compute limits:
\State \hspace{1em} $\mathrm{LL} = Q1 - 1.5 \cdot \mathrm{IQR}$ \hspace{1em} (Lower Limit)
\State \hspace{1em} $\mathrm{UL} = Q3 + 1.5 \cdot \mathrm{IQR}$ \hspace{1em} (Upper Limit)

\Statex
\For{\textbf{each data point} $x$}
  \State Compute anomaly score $s$
  \If{$s < Q1$ \textbf{or} $s > Q3$}
    \State Classify $x$ as \textbf{Anomalous} $\rightarrow D_a$
  \EndIf
  \If{$\mathrm{LL} \leq s \leq \mathrm{UL}$}
    \State Classify $x$ as \textbf{Normal} $\rightarrow D_n$
  \EndIf
\EndFor

\Statex
\State \textbf{Stage 2 of FAST-IDS ALG}
\State \textbf{Input:} $D_a$ (data classified as anomalies from Stage~1)
\State // $c$ = CNN feature maps, $l$ = LSTM hidden representation, $y_{pred}$ = class probabilities

\For{\textbf{each batch of sequences from} $D_a$}
    \State $c \gets \mathrm{ReLU}(\mathrm{Conv1D}_2(\mathrm{ReLU}(\mathrm{Conv1D}_1(D_a))))$
    \State $l \gets \mathrm{LSTM}(c)$
    \State $y_{pred} \gets \mathrm{Softmax}(W l + b)$
    \State Assign class label $\hat{y} \gets \arg\max(y_{pred})$
\EndFor

\end{algorithmic}
\end{algorithm}

\begin{algorithm}
\caption{Hybrid Model Compression}\label{alg:compression}
\small
\begin{algorithmic}[1]

\State \textbf{Input:} $w^{pre}$ = weights before compression, $\Theta$ = pre-trained full precision model
\State // $p$ = prune factor (0.4), $\mathcal{D}$ = calibration data, $E$ = fine-tuning epochs (10)
\State // $I_w$ = weight importance (L1 norm), $s$ = scale factor, $z$ = zero-point, $n$ = no. of bits to quantize, $w_\text{min}$, $w_\text{max}$ = min / max values in weight tensor, $x_\text{min}$ = minimum of the activation outputs
\State \textbf{Output:} Compressed model $\Theta^q$ with quantized weights $w^q$ and quantized activation $x^q$.

\For{each layer $o_l$ in $\Theta$}
    \State Compute weight importance: $I_w = \sum |w^{pre}|$ 
    \State Rank filters based on $I_w$
    \State Prune least important $p = 40\%$ filters $\;\; \to \;\; w^{pruned}$
\EndFor

\State Fine-tune the pruned model $\Theta^*$ on $\mathcal{D}$ for $E = 10$ epochs to obtain $w^{ft}$

\State Run calibration on $\mathcal{D}$ to determine $w_{\min}, w_{\max}$ and $x_{min}$

\For{each weight tensor $w^{ft}$ in $\Theta^*$}
    \State Compute scale factor: $s = \frac{w_{\max} - w_{\min}}{2^n - 1}$
    \State Compute zero-point: $z = \text{round}\!\left(-\frac{w_{\min}}{s}\right)$
    \State Quantize weights (symmetric):
    \[
       w^q = \text{clamp}\!\left(
              \text{round}\!\left(\frac{w^{ft}}{s}\right),
              -2^{n-1},\; 2^{n-1}-1
           \right)
    \]
\EndFor

\For{each activation $x$ }
    \State Quantize activations (asymmetric):
    \[
       x^q = \text{clamp}\!\left(
              \text{round}\!\left(\frac{x }{s}\right) + z,
              0,\; 2^n - 1
           \right)
    \]
\EndFor

\State Save the final compressed quantized model $\Theta^q$, $s$ and $z$ for dequantization. 

\end{algorithmic}
\end{algorithm}

\section{Experimental Setup}
\label{sec:simulation}
This section provides details of the system specification, dataset specification, and model hyperparameters used in this study. 

\subsection{System Specification}

The experiments were performed on an NVIDIA RTX A6000 GPU using a software environment that included Python 3.10.12, torch 2.1.1 and the VSCode IDE. The inference is carried out in NVIDIA RTX A6000, Jetson Nano (jetpack 4.6.6) and Google Colab CPU.

\subsection{Dataset Specification}
Table \ref{tab:datainstances} presents the instances used for training and testing in Stage 1. The anomalous data from Stage 1 undergoes labelling per class and sequence generation with a window size of 20, which resulted in 57,000 sequences. The process resulted in the following number of sequences per class: A(1) - 3804, A(2) - 3793, A(3) - 3821, A(4) - 3701, A(5) - 3623, A(6) - 3886, A(7) - 3662, A(8) - 3705, A(9) - 3727, A(10) - 3776, A(11) - 3870, A(12) - 3745, A(13) - 12564, A(14) - 12103, A(15) - 12370, A(16) - 16981, A(17) - 3876, A(18) - 8122 and A(19) - 7654. 
\begin{table}[h!]
\centering
\caption{Data Instances}
\resizebox{0.9\columnwidth}{!}{%
\begin{tabular}{cccc}
        \toprule
        S. No. & Instances & No. of Instances & Size \\
        \midrule
        1 & groundtruth / training & 30,50,501 & 1.3 gb \\
        2 & normal / testing & 1,18,78,433 & 6.1 gb \\
        3 & anomaly / testing & 52,47,975 & 2.8 gb \\
        \bottomrule
\end{tabular}%
}
\label{tab:datainstances}
\end{table}

\subsection{Training and Testing}

The training time for Stage 1 was 93,357.48 seconds, approximately 25.93 hours, whereas Stage 2 required 12,862.64 seconds, approximately 3.57 hours for modelling. In Stage 1, the model was trained for 30 epochs using the RMSProp optimizer with a learning rate of 0.0002. The gradient penalty lambda coefficient was set to 10, and the latent dimension was 100. In Stage 2, the model was trained for 100 epochs using the Adam optimizer with a learning rate of 0.0003.

\section{ Performance Evaluation and Analysis of Results}
\label{sec:results}
This section provides numerical and graphical results of the proposed FAST-IDS network, demonstrating its effectiveness against various vehicular attacks. Additionally, the analysis includes comparisons with baseline SOTA methods, detection of previously unseen attacks, and the model's performance when deployed in a resource-constrained environment.

\subsection{Evaluation and Analysis for Stage 1}
The models used in Stage 1 are as follows: 

\begin{itemize}
    \item BiGAN\_WGAN\_GP(5E\_4G\_4D) [M1] - BiGAN trained with Wasserstein's loss and Gradient Penalty, featuring a 5-layer CNN encoder, a 4-layer CNN generator, and a 4-layer CNN discriminator.
   \item  BiGAN\_WGAN\_GP(6E\_4G\_4D)  [M2] - BiGAN trained with Wasserstein's loss and Gradient Penalty, featuring a 6-layer CNN encoder, a 4-layer CNN generator, and a 4-layer CNN discriminator.
  \item BiGAN\_WGAN\_GP(5E\_5G\_5D)  [M3]  - BiGAN trained with Wasserstein's loss and Gradient Penalty, featuring a 5-layer CNN encoder, a 5-layer CNN generator, and a 5-layer CNN discriminator.
    \item BiGAN\_WGAN\_GP(4E\_4G\_4D)  [M4] - BiGAN trained with Wasserstein's loss and Gradient Penalty, featuring a 4-layer CNN encoder, a 4-layer CNN generator, and a 4-layer CNN discriminator.
    \item BiGAN\_WGAN(5E\_4G\_4D)  [M5] - BiGAN trained with Wasserstein's loss without Gradient Penalty.
    \item BiGAN\_BCE(5E\_4G\_4D)  [M6] - BiGAN trained with Binary Cross Entropy.
    \item BiGAN\_WGAN\_GP(5E\_4G\_4D)\_24h  [M7] - BiGAN trained for 24 hours files individually.
    
\end{itemize} 

\begin{figure}
    \centering
    \includegraphics[width=0.5\textwidth]{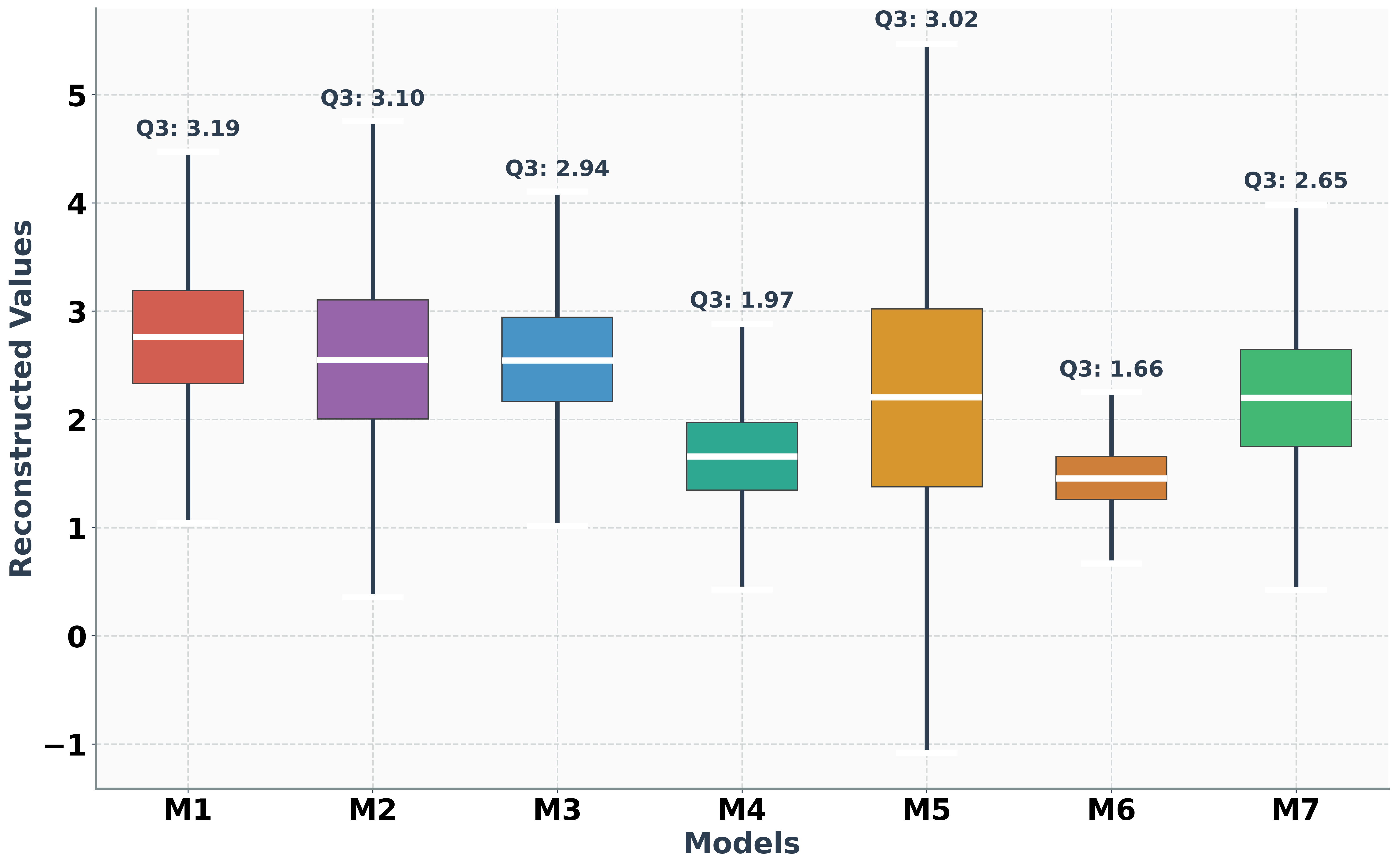}
    \caption{Whisker boxplots of reconstruction error for the seven BiGAN model variants used in the ablation study in Stage 1. Each boxplot illustrates the IQR ranges between 1st quartile (Q1) and the 3rd quartile (Q3) and extends upto the Lower (LL) and Upper Limits (UL).}
    \label{fig:comparison_iqr}
\end{figure}

The naming convention used for each of the models discussed above are closely related to the experiment configuration. Additionaly, we have also included a short name such as M1 to M7, which is used as an annotation in Figures (Fig. \ref{fig:comparison_iqr}.)
The recall values of Stage 1 models are illustrated in Table \ref{tab:rmse}, where Normal Recall represents the percentage of correctly identified normal instances, while Anomaly Recall indicates the percentage of correctly identified anomalous instances. 

\begin{table}[]
\centering
\caption{Recall values of Stage 1 models, highlighting the impact of architectural variations including number of layers utilized in encoder, decoder and generator and loss functions used on detection performance.}
\resizebox{0.9\columnwidth}{!}{%
\begin{tabular}{ccccc}
        \toprule
        S. No. & Model used & Normal\_recall & Anomaly\_recall  \\
        \midrule
        1 & BiGAN\_WGAN\_GP(5E\_4G\_4D)  &0.9967	&0.859\\
        2 & BiGAN\_WGAN\_GP(6E\_4G\_4D) & 0.9998 &0.5813\\
        3 & BiGAN\_WGAN\_GP(5E\_5G\_5D) & 0.996 &0.6595\\
        4 & BiGAN\_WGAN\_GP(4E\_4G\_4D) & 0.9643 &0.5331\\
        5 & BiGAN\_WGAN(5E\_4G\_4D) & 1	&0.2565 \\
        6 & BiGAN\_BCE(5E\_4G\_4D) & 0.8397&	0.816 \\
        7 & BiGAN\_WGAN\_GP(5E\_4G\_4D)\_24h & 0.95137&0.504612 \\
        \bottomrule

\end{tabular}%
\label{tab:rmse}
}
\end{table}

BiGAN\_WGAN\_GP(5E\_4G\_4D) achieves strong recall rates for normal and anomalous instances, benefiting from Wasserstein’s loss with Gradient Penalty and its 5-layer CNN encoder architecture. In contrast, other variants such as BiGAN\_WGAN\_GP(6E\_4G\_4D) and BiGAN\_WGAN\_GP(5E\_5G\_5D) exhibit underfitting to anomalous data points, indicating that increasing the number of layers may negatively impact model performance. While BiGAN\_WGAN\_GP(4E\_4G\_4D)  performs well on normal instances but struggles with anomalies, suggested that its 4-layer CNN encoder lacks the complexity of the 5-layer variant. BiGAN\_WGAN(5E\_4G\_4D) achieves perfect recall for normal instances but overfits anomalies, reflecting the absence of Gradient Penalty. BiGAN\_BCE(5E\_4G\_4D) maintains strong performance across both instance types, benefiting from Binary Cross Entropy’s reduced sensitivity to extreme values. Meanwhile, BiGAN\_WGAN\_GP(5E\_4G\_4D)\_24h adapts to 24-hour data imbalance but underperforms on anomalies. Overall, BiGAN\_WGAN\_GP(5E\_4G\_4D)  offers the best recall for both normal and anomalous instances, demonstrating robustness in anomaly detection, highlighting the importance of Gradient Penalty and complex CNN architectures in improving anomaly detection.

To further analyze the BiGAN variants' behaviour, we plotted the whisker boxplot in Fig. \ref{fig:comparison_iqr}. The distributions are based on IQR ranges, highlighting the variability of each model's reconstruction error. The BiGAN\_WGAN(5E\_4G\_4D)  [M5] variant with broader reconstruction error, showcase poor performance, while model BiGAN\_WGAN\_GP(5E\_4G\_4D)  [M1], with tighter distributions and compact IQR, illustrate superiority in input reconstruction and anomaly detection. But as the plot becomes more narrower, the the model also struggles to detect anomaly (BiGAN\_WGAN\_GP(4E\_4G\_4D) [M4]) and normal data (BiGAN\_BCE(5E\_4G\_4D) [M6]).

\subsection{Evaluation and Analysis for Stage 2}

The models used in Stage 2 are as follows:
\begin{itemize}
\item  CNN\_LSTM\_CE  - The hybrid model, consisting of 1 CNN and 1 LSTM, was updated using cross-entropy loss.
\item 2CNN\_LSTM\_CE - The hybrid model, consisting of 2 CNN and 1 LSTM, was updated using cross-entropy loss.
\item  2CNN\_BiLSTM\_CE - The hybrid model, consisting of 2 CNN and 1 Bidirectional LSTM, was updated using cross-entropy loss.
\item 2CNN\_LSTM\_LSmooth - The hybrid model, consisting of 2 CNN and 1 LSTM, was updated using label smoothing Loss.
\item 2CNN\_LSTM\_FL - The hybrid model, consisting of 2 CNN and 1 LSTM, was updated using focal loss.

\end{itemize}
\begin{table}[t]
\centering
\caption{Overall Performance Metrics Across Stage 2 Models, highlighting the impact of number of layers and loss function in the hybrid CNN-LSTM model.}
\label{tab:stage2mterics}
\small
\renewcommand{\arraystretch}{1.2}
\begin{tabular}{lcccc}
\toprule
\textbf{Model} & \textbf{Acc} & \textbf{Precision} & \textbf{Recall} & \textbf{F1-score} \\
\midrule
CNN\_LSTM\_CE         & 97.81 & 78.55 & 78.07 & 78.05 \\
2CNN\_LSTM\_CE        & 97.80 & 78.75 & 78.01 & 78.17 \\
2CNN\_BiLSTM\_CE      & 97.81 & 78.55 & 78.07 & 78.05 \\
2CNN\_LSTM\_LSmooth   & 97.88 & 79.42 & 78.81 & 78.21 \\
2CNN\_LSTM\_FL        & 97.81 & 78.72 & 78.14 & 78.15 \\
\bottomrule
\end{tabular}
\end{table}

The results presented in Tables \ref{tab:stage2mterics} illustrate the performance metrics for Stage 2 models.

\begin{table*}[t]
\centering
\caption{Summarized results illustrating the effectiveness of our Two-Stage Intrusion Detection Framework.}
\label{tab:stage_results}
\small
\renewcommand{\arraystretch}{1.2}
\begin{tabular}{l l c c c c}
\toprule
\textbf{Stage} & \textbf{Model} & \textbf{Accuracy (\%)} & \textbf{Precision (\%)} & \textbf{Recall (\%)} & \textbf{F1-score (\%)} \\
\midrule
Stage 1 & BiGAN\_WGAN\_GP(5E\_4G\_4D) & 95.44   & 100    & 92.79 & 96.12 \\
Stage 2 & 2CNN\_LSTM\_LSmooth         & 97.88 & 79.42 & 78.81 & 78.21 \\
\bottomrule
\end{tabular}
\end{table*}

\begin{figure}
    \centering
    \includegraphics[width=1\linewidth]{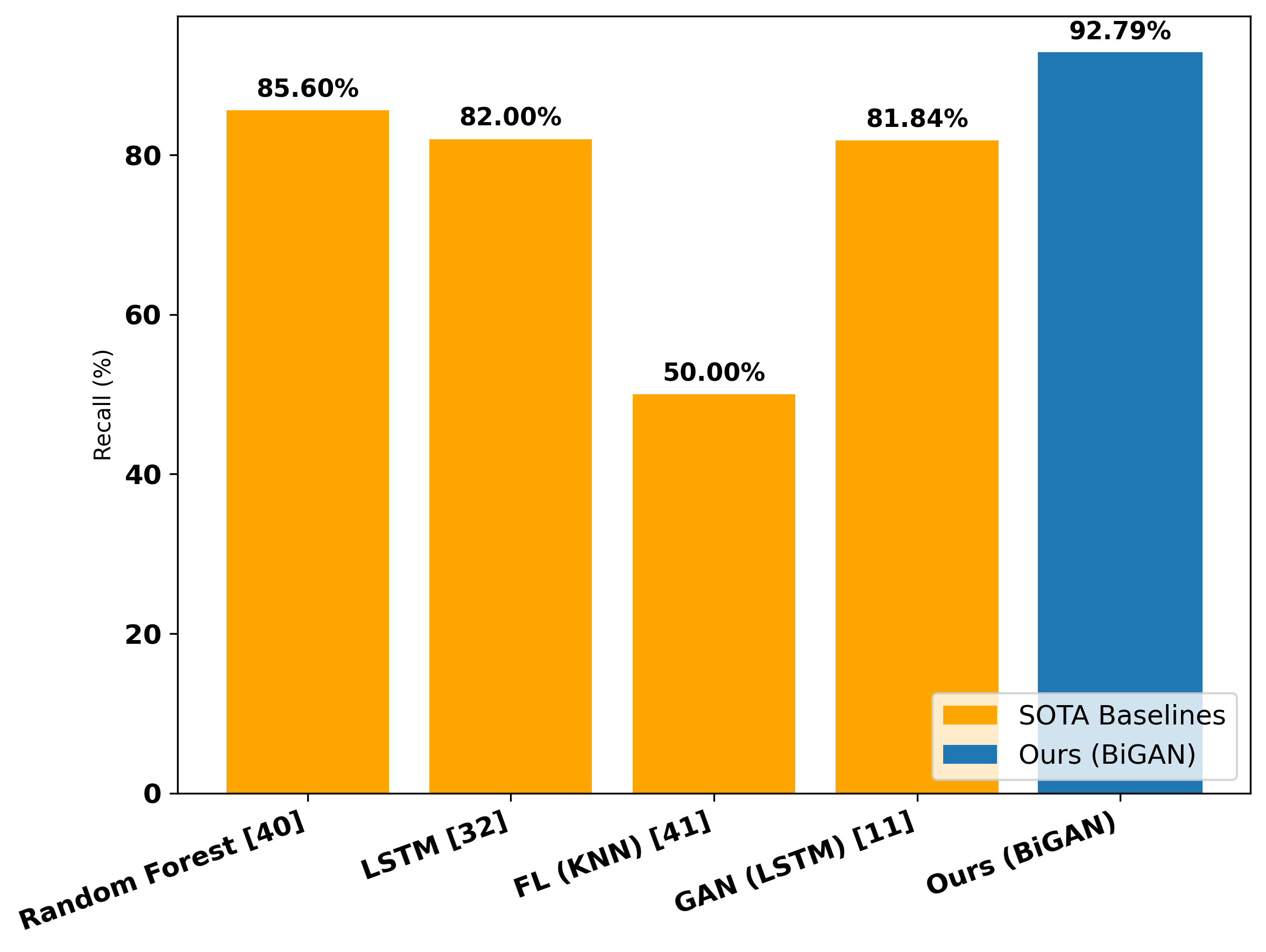}
    \caption{Graph illustrating the performance of FAST-IDS (Stage 1) with respect to other SOTA baselines in terms of recall percentages.}
    \label{fig:SOTA_STAGE1}
\end{figure}

\begin{figure}
    \centering
    \includegraphics[width=1.0\linewidth]{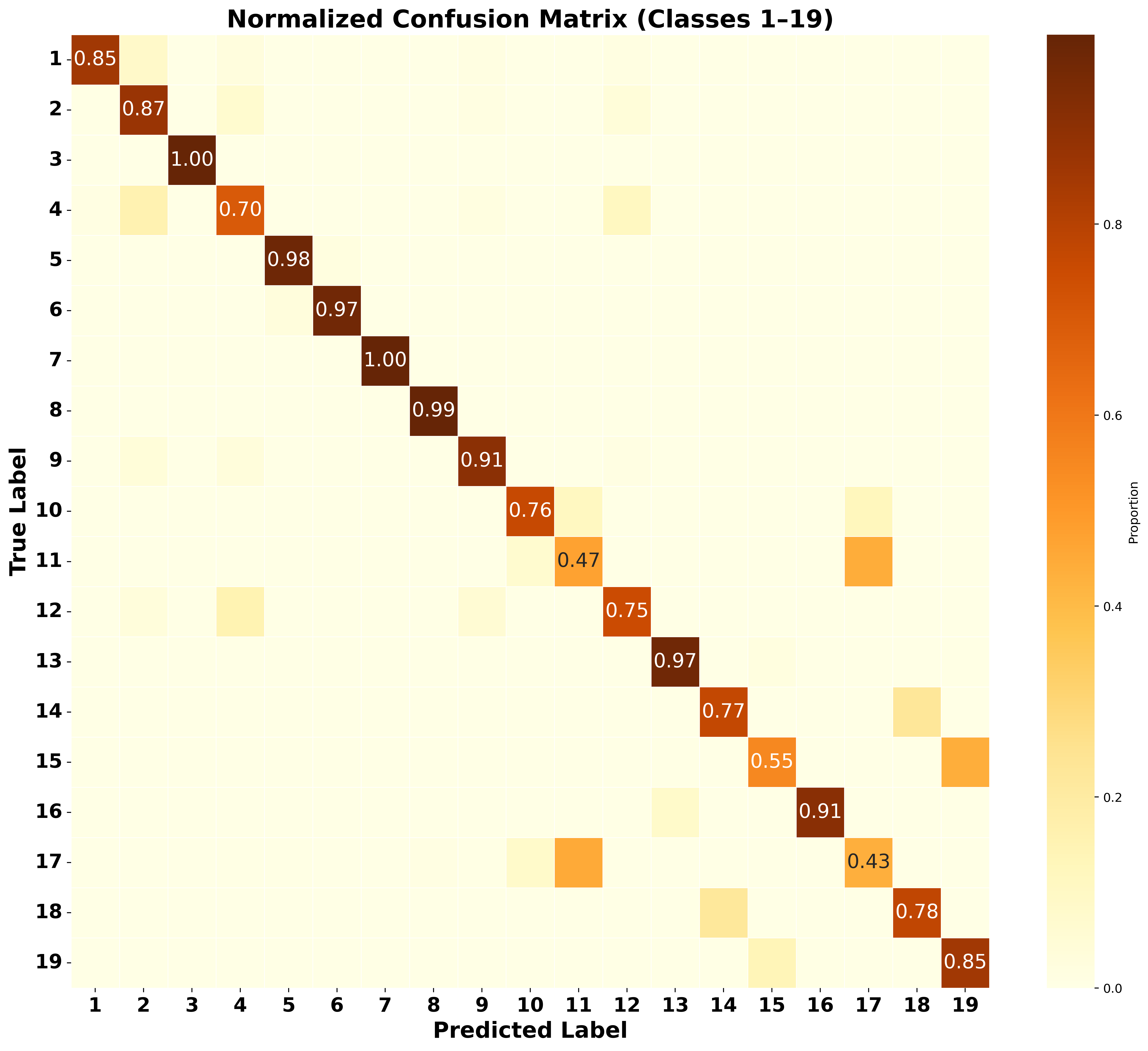}
    \caption{The confusion matrix visualized as the heat map of the Stage 2 Hybrid CNN-LSTM model evaluated across 19 attack classes.}
    \label{fig:confusion_matrix}
\end{figure}

\begin{figure}[h!]
    \centering
    \begin{subfigure}[b]{0.4\textwidth}
        \centering
        \includegraphics[width=\textwidth]{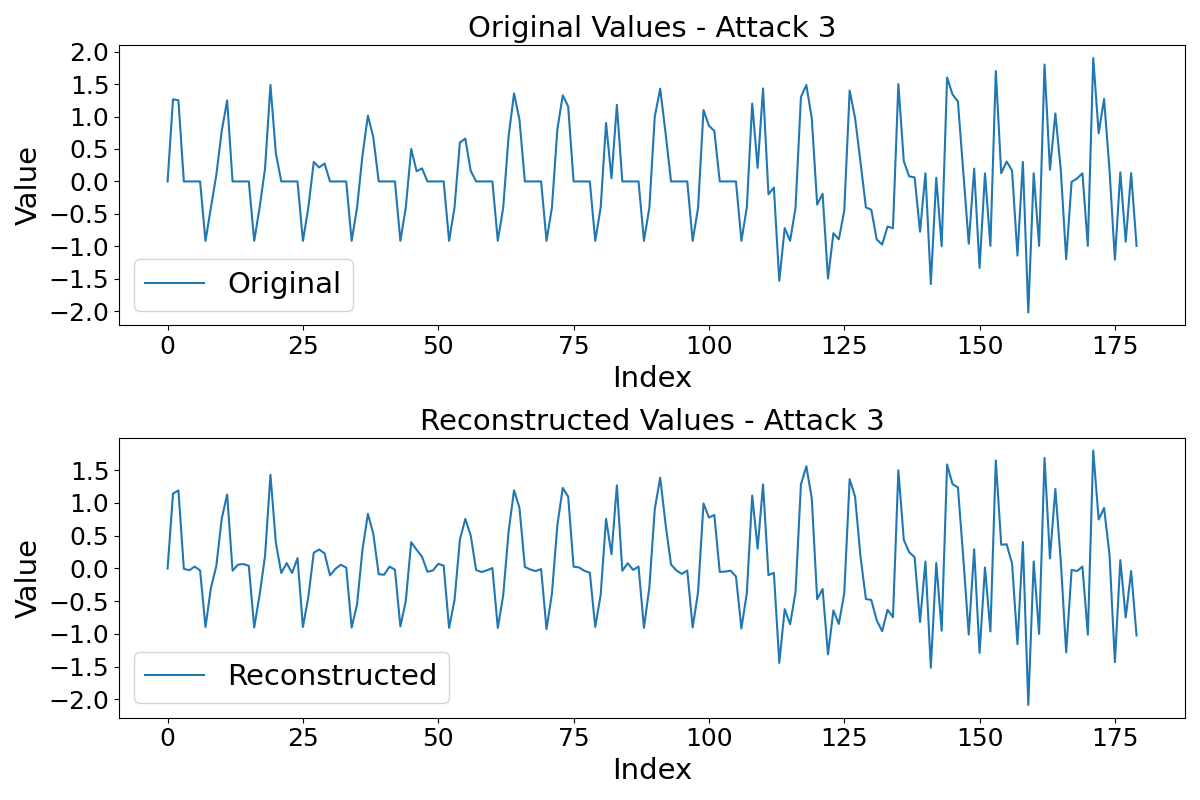}
        \caption{Original and Reconstructed Plots for Attack 3.}
        \label{fig:rec3}
    \end{subfigure}
    
    \begin{subfigure}[b]{0.4\textwidth}
        \centering
        \includegraphics[width=\textwidth]{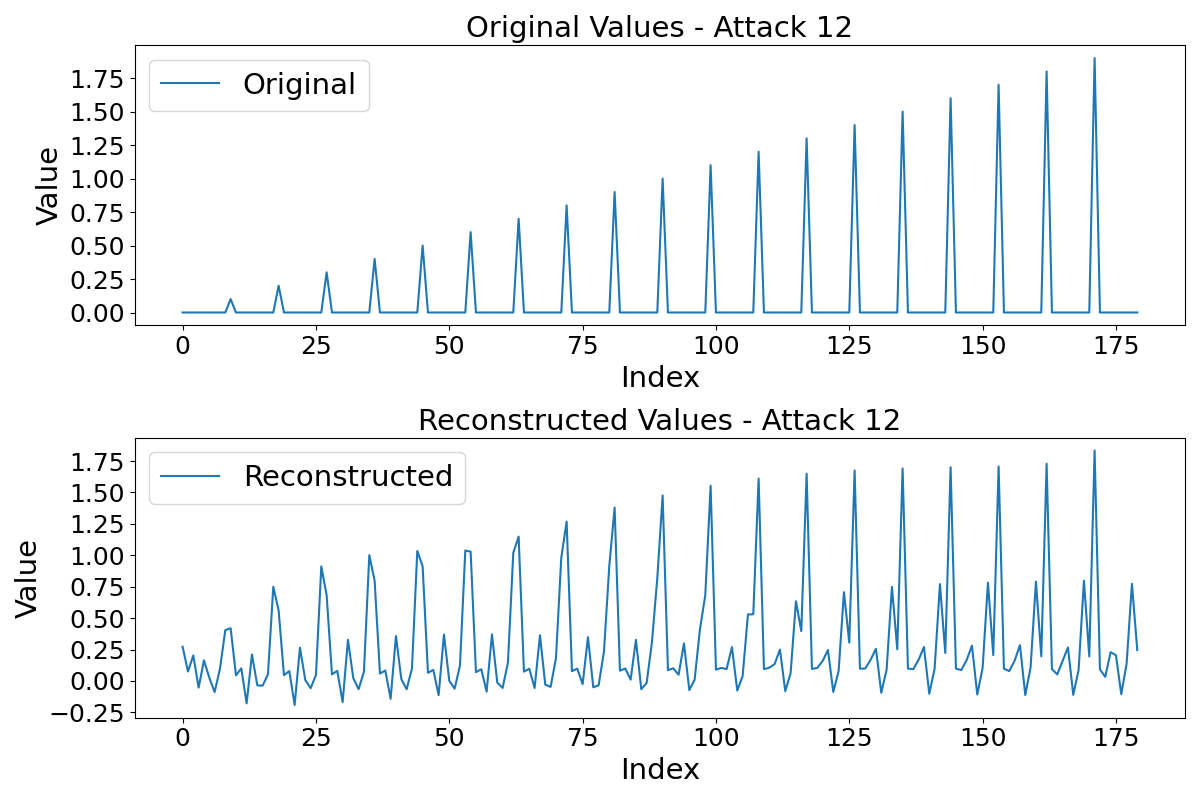}
        \caption{Original and Reconstructed Plots for Attack 12.}
        \label{fig:rec12}
    \end{subfigure}
    \caption{Comparison of Original and Reconstructed Plots}
    \label{fig:comparison}
\end{figure}

\begin{table}[t]
\centering
\caption{Comparison of Recall Rates for Various Attack Classes with SOTA for Stage 2}
\label{tab:recall-comparison}
\small
\resizebox{0.95\columnwidth}{!}{%
\begin{tabular}{@{}c|ccccc@{}}
\toprule
\textbf{Attack Class} & \textbf{Our Model} & \textbf{\cite{chougule2022multibranch}} & \textbf{\cite{slama2022comparative}} & \textbf{\cite{hsu2021deep}} & \textbf{\cite{slama2023one}} \\
\midrule
A(1)  & 82.50 & 42.5 & 82.34 & 54.12 & 65 \\
A(2)  & 78.33 & 5.5  & 74.67 & 99.96 & 69 \\
A(3)  & 100   & 100  & 99.30 & 99.82 & 61 \\
A(4)  & 60.83 & 8    & 99.36 & 99.20 & 71 \\
A(5)  & 97.83 & 94.5 & 81.81 & 99.88 & 59 \\
A(6)  & 97.17 & 42   & 3.81  & 100   & 61 \\
A(7)  & 100   & 100  & 99.48 & 98.73 & 49 \\
A(8)  & 100   & 100  & 98.55 & 100   & 61 \\
A(9)  & 85.83 & 9    & 6.53  & 99.70 & 71 \\
A(10) & 79.17 & 98.5 & 6.81  & 99.73 & 58 \\
A(11) & 53.83 & 98.5 & 99.92 & 78.12 & 62 \\
A(12) & 65.17 & 8.5  & 99.90 & 99.97 & 56 \\
A(13) & 96.17 & 98   & 78.25 & 94.91 & 56 \\
A(14) & 78.67 & 100  & 78.88 & 99.98 & 50 \\
A(15) & 56.83 & 99.5 & 60.04 & 99.93 & 56 \\
A(16) & 98.50 & 98.5 & 2.78  & 100   & 62 \\
A(17) & 43.17 & 97.5 & 53.26 & 100   & 58 \\
A(18) & 91.83 & 100  & 79.58 & 88.23 & 62 \\
A(19) & 86.83 & 99   & 76.93 & 99.67 & 63 \\
\bottomrule
\end{tabular}
}
\end{table}

\textcolor{black}{The choice of hybrid CNN-LSTM models is justified by their strong performance, as evidenced in \cite{alladi2021artificial}, \cite{chougule2024hybridsecnet} and \cite{alladi2021deepadv}.
Among the models presented in Table \ref{tab:stage2mterics}, CNN\_LSTM\_CE underperforms in performance compared to other models, while 2CNN\_LSTM\_CE achieves the better accuracy and balanced performance due to an additional CNN layer capturing more complex features. The bidirectional variant (2CNN\_BiLSTM\_CE) performs similarly to 2CNN\_LSTM\_CE, showing minimal improvement over 2CNN\_LSTM\_CE in accuracy and recall. 2CNN\_LSTM\_LSmooth demonstrates comparatively the best performance, indicating that label smoothing loss effectively balances these metrics and slightly outperforms the focal loss model (2CNN\_LSTM\_FL). The results emphasize the role of model architecture and loss functions in multiclass classification, with 2CNN\_LSTM\_LSmooth delivering balanced performance. Since 2CNN\_LSTM\_LSmooth yielded the best results, we experimented with increasing the layers of CNN and LSTM for this variant. While this maintained performance similar to 2CNN\_LSTM\_LSmooth, it resulted in a significant increase in computational time. The confusion matrix presented in Fig. \ref{fig:confusion_matrix} provides a detailed view of the classification performed by the 2CNN\_LSTM\_LSmooth model, demonstrating the correctly and the wrongly classified samples per class.}

Our study observed that only limited studies have performed multi-stage intrusion detection in CAVs utilizing the VeReMi extension dataset \cite{kamel2020veremi}. For Stage 1, we have compared our work with existing methodologies implementing binary classification \cite{drenyovszki2024development,
kushardianto2024vehicular,
moulahi2023privacy} or anomaly detection systems \cite{devika2024vadgan}. In contrast, for Stage 2, we benchmarked against studies that focused on multi-class classification \cite{chougule2022multibranch,slama2022comparative,
hsu2021deep,slama2023one}.

The results corresponding to the best performing models in Stage 1 and 2 are presented in Table \ref{tab:stage_results}.

\begin{table*}[ht] 
\centering
\caption{Evaluation of Model Compression Techniques Across Different Environments}
\large 
\renewcommand{\arraystretch}{1.3} 
\setlength{\tabcolsep}{8pt} 
\resizebox{0.9\textwidth}{!}{ 
\begin{tabular}{m{3.5cm} m{3.5cm} m{3.5cm} m{3.5cm} m{3.5cm} m{3.5cm}} 
        \toprule
       \multirow{2}{*}{\textbf{Environment}} & \multirow{2}{*}{\textbf{Inference Time (s)}} & \multirow{2}{*}{\textbf{M/y Utilization (MiB)}} & \multicolumn{3}{c}{\textbf{Compression Statistics}} \\ 
       \cmidrule(lr){4-6} 
       & & & \textbf{Model Reduction (\%)} & \textbf{Performance Loss (\%)} & \textbf{Reduced FLOPs / MMACs (\%)} \\ 
       \midrule
       \textbf{RTX A6000 GPU} & 2.2765  & 3428.21  & \multirow{3}{*}{77.2} & \multirow{3}{*}{5} & \multirow{3}{*}{34.17 / 54.15} \\ 
       \textbf{Google Colab CPU} & 7.938  & 1636.52  &  &  &  \\ 
       \textbf{Jetson Nano} & 102.62 & 3353.34 &  &  &  \\ 
        \bottomrule
\end{tabular}%
}
\label{tab:eff}
\end{table*}

\subsection{Stage 1 Comparison with SOTA} An extensive comparison with baseline models \cite{drenyovszki2024development,
kushardianto2024vehicular,
moulahi2023privacy,devika2024vadgan} are illustrated in Fig. \ref{fig:SOTA_STAGE1}.  Our model demonstrates superior overall performance compared to the SOTA, with a recall percentage of 92.785\%. Compared to Random Forest utilized in \cite{drenyovszki2024development}, LSTM in \cite{kushardianto2024vehicular}, Federated Learning with KNN utilized in \cite{moulahi2023privacy}, and GAN with LSTM utilized in \cite{devika2024vadgan}, our model achieved a relative improvement of 30.25\%, clearly demonstrating the superiority of FAST-IDS in detecting anomalous behaviour. This relative improvement is due to the integration of BiGAN with WGAN features, which stabilized training of the model. Unlike traditional models such as KNN and Random Forest, which provided limited representation learning and standard GAN prone to model collapse, our model succeeded in capturing both temporal and spatial structures in the data.

\subsection{Stage 2 Comparison with SOTA}

Stage 2 is evaluated against single-stage classifiers \cite{chougule2022multibranch,slama2022comparative,
hsu2021deep,slama2023one} to demonstrate the effectiveness of our multi-stage model and the comparisons are illustrated in Table \ref{tab:recall-comparison}. Our results are comparable in 11 attack types in both \cite{chougule2022multibranch} and \cite{slama2022comparative} individually - Constant position (A1), Constant position offset (A2), Random position (A3), Random position offset (A4), Constant speed (A5), Constant speed offset (A6), Random speed (A7), Random speed offset (A8), Eventual Stop (A9), Delayed messages (A12) and Data replay Sybil (A16) \cite{chougule2022multibranch}. While \cite{chougule2022multibranch} employed a CNN, we utilized Hybrid CNN-LSTM models. This difference might explain why \cite{chougule2022multibranch} yielded very low recall values for Constant position offset (A2), Random position offset (A4), Eventual Stop (A9), and Delayed messages (A12). On the other hand, \cite{hsu2021deep} developed an integrated approach combining CNN, LSTM and Support Vector Machine (SVM) and demonstrated exceptional results, but it lacks generalization capability. In comparison with \cite{slama2023one} which uses a tree-based classifier, our model demonstrates a clear advantage in outperforming in 16 attacks.

\subsection{Detection of Previously Unseen Attacks}

Stage 1 employs BiGAN for detecting unseen data through its reconstructive properties. In Stage 2, the dataset was split into known (A0–A9) and unknown (A10–A19) attacks. The model was trained and validated on known attacks and tested on known and unknown attacks. A reconstructive layer was added after LSTM, enabling the model to learn input patterns effectively. The Mean Squared Error (MSE) between the input and the reconstructed data was used as the reconstruction error. We have employed a percentile-based thresholding strategy to address the dynamic and evolving nature of attacks. Here, the 91st percentile of the reconstruction errors from the validation set was selected as the decision boundary. Various threshold ranges between 80 and 93 were evaluated, while 91 yielded the best tradeoff between known and unknown classes. The accuracies within unseen data are as follows: Class 10: 97.83\%, Class 11: 82.33\%, Class 12: 50.17\%, Class 13: 54.67\%, Class 14: 100.00\%, Class 15: 100.00\%, Class 16: 81.17\%, Class 17: 84.50\%, Class 18: 100.00\%, and Class 19: 99.83\%. As shown in Fig. \ref{fig:rec3} and \ref{fig:rec12}, the reconstructed plot for A(3) (known attack) closely matches the original, confirming accurate reconstruction, whereas A(12) (unknown attack) deviates significantly, validating its classification as an anomaly.

\subsection{Computational Efficiency}

\textcolor{black}{To enhance efficiency and manage computational demands, we evaluated the model using hybrid data compression techniques, incorporating L1-based structured pruning followed by static quantization. We achieved a 54.4\% reduction in model size through pruning and  77.2\% reduction through quantization, leading to 34.17\% reduction in FLOPs and a 54.15\% reduction in MACs,  thereby significantly enhancing computational efficiency.
This optimization resulted in approximately 50.05\% reduction in inference time with 5\% performance loss, as presented in Table \ref{tab:eff}, demonstrating significant computational efficiency improvements.} Our model achieved attack detection within 0.195 seconds (per vehicle) on Jetson Nano, highlighting its efficiency for real-time deployment in resource-constrained environments. \textcolor{black}{As part of the ablation study, we also evaluated random filter pruning, achieving 47.36\% model compression but with a 16\% performance loss. Additionally, L1 unstructural pruning led to a 40.59\% model size reduction but achieved only 1.5× faster inference, highlighting the effectiveness of our proposed hybrid approach.}

\subsection{Comparison of Run Time Results with SOTA}

The inference time for data setup and prediction is critical for real-time IDS performance. During inference, 10,000 sequences corresponding to 526 vehicles were considered. As shown in Table \ref{tab:inference_time}, our compressed model achieves faster inference than \cite{chougule2022multibranch} utilizing a reconstruction-based CNN model, \cite{alladi2023deep} using a 4-layer LSTM, and \cite{alladi2021deepadv} using a hybrid CNN-BiLSTM, significantly improving computational efficiency. This enhancement makes the model well-suited for real-time intrusion detection in autonomous vehicles, enabling fast threat identification and response in dynamic vehicular networks.

\begin{table}[H]
    \centering
    \caption{Inference Time Comparison on Jetson Nano with SOTA}
    \renewcommand{\arraystretch}{1.2} 
    \setlength{\tabcolsep}{6pt}
    \resizebox{0.5\textwidth}{!}{
    \begin{tabular}{c|c|c|c|c}
        \toprule
        \textbf{Reference} & \textbf{Environment} & \textbf{Data Setup} & \textbf{Prediction} & \textbf{Inference Time (ms)} \\
        \midrule
        \cite{chougule2022multibranch} & \multirow{3}{*}{Jetson Nano} & 0.07 & 511.82 & 511.89 \\
        \cite{alladi2023deep} &  & 287.74 & 0.33 & 288.07 \\
        \cite{alladi2021deepadv} &  &  370.92 & 0.395 & 371.32 \\
        Ours &  & 0.003 & 194.997 & \textbf{195} \\
        \bottomrule
    \end{tabular}
    }
    \label{tab:inference_time}
\end{table}

\section{Conclusion}
\label{sec:conc}
 
This paper presents a two-stage intrusion detection framework for Connected and Autonomous Vehicles (CAVs), optimized through hybrid model compression for enhanced computational efficiency and real-time deployment. The compression technique combines structural pruning using the L1 norm and static quantization, utilizing a min-max observer for calibration. This reduces model size by 77.2\% and achieves approximately 50.05\% reduction in inference time, enabling per-vehicle attack detection on Jetson Nano in 0.195 ms, making it suitable for resource-constrained Intelligent Transportation Systems (ITS). The proposed two-stage architecture combines a Bidirectional Generative Adversarial Network (BiGAN) with Wasserstein loss, gradient penalty, and a deep CNN architecture leveraging reconstruction error using Mean Squared Error (MSE) and Mahalanobis distance, with a hybrid CNN-LSTM in Stage 2 trained with label smoothing loss. Together, both stages enable the detection of previously unseen attacks. At the same time, hybrid model compression ensures efficient, accurate, and real-time attack detection in CAV networks, making it highly adaptable for real-world ITS applications. Our two-stage model explicitly addresses the uncertainties in IDS, such as previously unseen attack types, changing traffic scenarios, and unpredictable behaviour of attackers, through the BiGAN anomaly detector and the Hybrid CNN-LSTM intrusion detector. Although the hybrid compression technique may introduce additional uncertainties, our framework achieved comparable detection rates and faster inference over baseline models.

Future work will focus on developing more lightweight architectures, advancing model compression techniques, and exploring Federated Learning for decentralized training to enhance data privacy and reduce communication overhead in large-scale vehicular networks. Also, we plan to explore adaptive feature elimination and redundancy-aware mechanisms to deal with computational efficiency, and an online thresholding mechanism to deal with anomaly detection.

\bibliographystyle{IEEEtranN}
{\footnotesize
\bibliography{ref}}

@article{alladi2021artificial,
  title={Artificial intelligence (AI)-empowered intrusion detection architecture for the internet of vehicles},
  author={Alladi, Tejasvi and Kohli, Varun and Chamola, Vinay and Yu, F Richard and Guizani, Mohsen},
  journal={IEEE Wireless Communications},
  volume={28},
  number={3},
  pages={144--149},
  year={2021},
  publisher={IEEE}
}

@inproceedings{amanullah2022burst,
  title={BurST-ADMA: Towards an Australian dataset for misbehaviour detection in the internet of vehicles},
  author={Amanullah, Mohamed Ahzam and Chhetri, Mohan Baruwal and Loke, Seng W and Doss, Robin},
  booktitle={Proceedings of the IEEE International Conference on Pervasive Computing and Communications Workshops and other Affiliated Events (PerCom Workshops)},
  pages={624--629},
  year={2022}
}

@ARTICLE{fan2024auto,
  author={Fan, Chunyang and Cui, Jie and Jin, Hulin and Zhong, Hong and Bolodurina, Irina and He, Debiao},
  journal={IEEE Transactions on Vehicular Technology}, 
  title={Auto-Updating Intrusion Detection System for Vehicular Network: A Deep Learning Approach Based on Cloud-Edge-Vehicle Collaboration}, 
  year={2024},
  volume={73},
  number={10},
  pages={15372-15384},
  doi={10.1109/TVT.2024.3399219}}

@article{sharma2020machine,
  title={A machine-learning-based data-centric misbehavior detection model for internet of vehicles},
  author={Sharma, Prinkle and Liu, Hong},
  journal={IEEE Internet of Things Journal},
  volume={8},
  number={6},
  pages={4991--4999},
  year={2020},
  publisher={IEEE}
}

@inproceedings{sharma2019attacks,
  title={Attacks on machine learning: Adversarial examples in connected and autonomous vehicles},
  author={Sharma, Prinkle and Austin, David and Liu, Hong},
  booktitle={Proceedings of the IEEE International Symposium on Technologies for Homeland Security (HST)},
  pages={1--7},
  year={2019}
}

@article{alladi2023deep,
  title={A deep learning based misbehavior classification scheme for intrusion detection in cooperative intelligent transportation systems},
  author={Alladi, Tejasvi and Kohli, Varun and Chamola, Vinay and Yu, F Richard},
  journal={Digital Communications and Networks},
  volume={9},
  number={5},
  pages={1113--1122},
  year={2023},
  publisher={Elsevier}
}

@ARTICLE{devika2024vadgan,
  author={S, Devika and Shrivastava, Rishi Rakesh and Narang, Pratik and Alladi, Tejasvi and Yu, F. Richard},
  journal={IEEE Transactions on Vehicular Technology}, 
  title={VADGAN: An Unsupervised GAN Framework for Enhanced Anomaly Detection in Connected and Autonomous Vehicles}, 
  year={2024},
  volume={73},
  number={9},
  pages={12458-12467},
  doi={10.1109/TVT.2024.3388591}}

@article{xie2021threat,
  title={Threat analysis for automotive CAN networks: A GAN model-based intrusion detection technique},
  author={Xie, Guoqi and Yang, Laurence T and Yang, Yuanda and Luo, Haibo and Li, Renfa and Alazab, Mamoun},
  journal={IEEE Transactions on Intelligent Transportation Systems},
  volume={22},
  number={7},
  pages={4467--4477},
  year={2021},
  publisher={IEEE}
}

@article{zhao2022can,
  title={CAN bus intrusion detection based on auxiliary classifier GAN and out-of-distribution detection},
  author={Zhao, Qingling and Chen, Mingqiang and Gu, Zonghua and Luan, Siyu and Zeng, Haibo and Chakrabory, Samarjit},
  journal={ACM Transactions on Embedded Computing Systems (TECS)},
  volume={21},
  number={4},
  pages={1--30},
  year={2022},
  publisher={ACM New York, NY}
}

@article{chougule2022multibranch,
  title={Multibranch reconstruction error (mbre) intrusion detection architecture for intelligent edge-based policing in vehicular ad-hoc networks},
  author={Chougule, Amit and Kohli, Varun and Chamola, Vinay and Yu, Fei Richard},
  journal={IEEE Transactions on Intelligent Transportation Systems},
  volume={24},
  number={11},
  pages={13068--13077},
  year={2022},
  publisher={IEEE}
}

@article{festag2014cooperative,
  title={Cooperative intelligent transport systems standards in Europe},
  author={Festag, Andreas},
  journal={IEEE Communications Magazine},
  volume={52},
  number={12},
  pages={166--172},
  year={2014},
  publisher={IEEE}
}

@article{shladover2018connected,
  title={Connected and automated vehicle systems: Introduction and overview},
  author={Shladover, Steven E},
  journal={Journal of Intelligent Transportation Systems},
  volume={22},
  number={3},
  pages={190--200},
  year={2018},
  publisher={Taylor \& Francis}
}

@inproceedings{kamel2020veremi,
  title={Veremi extension: A dataset for comparable evaluation of misbehavior detection in vanets},
  author={Kamel, Joseph and Wolf, Michael and Van Der Hei, Rens W and Kaiser, Arnaud and Urien, Pascal and Kargl, Frank},
  booktitle={Proceedings of the ICC 2020-2020 IEEE International Conference on Communications (ICC)},
  pages={1--6},
  year={2020}
}

@article{donahue2016adversarial,
  title={Adversarial feature learning},
  author={Donahue, Jeff and Kr{\"a}henb{\"u}hl, Philipp and Darrell, Trevor},
  journal={arXiv preprint arXiv:1605.09782},
  year={2016}
}

@article{xu2018joint,
  title={Joint reconstruction and anomaly detection from compressive hyperspectral images using Mahalanobis distance-regularized tensor RPCA},
  author={Xu, Yang and Wu, Zebin and Chanussot, Jocelyn and Wei, Zhihui},
  journal={IEEE Transactions on Geoscience and Remote Sensing},
  volume={56},
  number={5},
  pages={2919--2930},
  year={2018},
  publisher={IEEE}
}

@inproceedings{lee2008multi,
  title={Multi-stage intrusion detection system using hidden markov model algorithm},
  author={Lee, Do-hyeon and Kim, Doo-young and Jung, Jae-il},
  booktitle={Proceedings of the International Conference on Information Science and Security (ICISS 2008)},
  pages={72--77},
  year={2008}
}

@article{zhou2021detecting,
  title={Detecting multi-stage attacks using sequence-to-sequence model},
  author={Zhou, Peng and Zhou, Gongyan and Wu, Dakui and Fei, Minrui},
  journal={Computers \& Security},
  volume={105},
  pages={102203},
  year={2021},
  publisher={Elsevier}
}

@inproceedings{van2018veremi,
  title={Veremi: A dataset for comparable evaluation of misbehavior detection in vanets},
  author={Van Der Heijden, Rens W and Lukaseder, Thomas and Kargl, Frank},
  booktitle={Security and Privacy in Communication Networks: 14th International Conference, SecureComm 2018, Singapore, Singapore, August 8-10, 2018, Proceedings, Part I},
  pages={318--337},
  year={2018},
  organization={Springer}
}

@article{ullah2022hdl,
  title={HDL-IDS: a hybrid deep learning architecture for intrusion detection in the Internet of Vehicles},
  author={Ullah, Safi and Khan, Muazzam A and Ahmad, Jawad and Jamal, Sajjad Shaukat and e Huma, Zil and Hassan, Muhammad Tahir and Pitropakis, Nikolaos and Arshad and Buchanan, William J},
  journal={Sensors},
  volume={22},
  number={4},
  pages={1340},
  year={2022},
  publisher={MDPI}
}

@inproceedings{slama2022comparative,
  title={Comparative study of misbehavior detection system for classifying misbehaviors on VANET},
  author={Slama, Omessaad and Alaya, Bechir and Zidi, Salah and Tarhouni, Mounira},
  booktitle={2022 8th International Conference on Control, Decision and Information Technologies (CoDIT)},
  volume={1},
  pages={243--248},
  year={2022},
  organization={IEEE}
}

@inproceedings{drenyovszki2024development,
  title={Development of an Attack Detection Module for Vehicular Ad-hoc Networks},
  author={Drenyovszki, Rajmund and Johany{\'a}k, Zsolt Csaba},
  booktitle={2024 IEEE 18th International Symposium on Applied Computational Intelligence and Informatics (SACI)},
  pages={000255--000260},
  year={2024},
  organization={IEEE}
}

@ARTICLE{chougule2024hybridsecnet,
  author={Chougule, Amit and Kulkarni, Ishan and Alladi, Tejasvi and Chamola, Vinay and Yu, Fei Richard},
  journal={IEEE Transactions on Vehicular Technology}, 
  title={HybridSecNet: In-Vehicle Security on Controller Area Networks Through a Hybrid Two-Step LSTM-CNN Model}, 
  year={2024},
  volume={73},
  number={10},
  pages={14580-14591},
  doi={10.1109/TVT.2024.3413849}}

@article{kumar2024survey,
  title={A survey on the blockchain techniques for the Internet of Vehicles security},
  author={Kumar, Sathish and Velliangiri, Sarveshwaran and Karthikeyan, Periyasami and Kumari, Saru and Kumar, Sachin and Khan, Muhammad Khurram},
  journal={Transactions on Emerging Telecommunications Technologies},
  volume={35},
  number={4},
  pages={e4317},
  year={2024},
  publisher={Wiley Online Library}
}

@article{ismail2024designing,
  title={Designing anonymous key agreement scheme for secure vehicular ad-hoc networks},
  author={Ismail, Md and Chatterjee, Santanu and Sing, Jamuna Kanta and Kumari, Saru and Rodrigues, Joel JPC},
  journal={IEEE Transactions on Intelligent Transportation Systems},
  year={2024},
  publisher={IEEE}
}

@misc{kumari2023safety,
  title={Safety and security of autonomous vehicles},
  author={Kumari, Saru and Xiong, Hu and Khoukhi, Lyes and Rodrigues, Joel JPC},
  journal={Transactions on Emerging Telecommunications Technologies},
  volume={34},
  number={11},
  pages={e4880},
  year={2023},
  publisher={John Wiley \& Sons, Ltd. Chichester, UK}
}

@article{chen2023provably,
  title={A provably secure key transfer protocol for the fog-enabled Social Internet of Vehicles based on a confidential computing environment},
  author={Chen, Chien-Ming and Li, Zhen and Kumari, Saru and Srivastava, Gautam and Lakshmanna, Kuruva and Gadekallu, Thippa Reddy},
  journal={Vehicular Communications},
  volume={39},
  pages={100567},
  year={2023},
  publisher={Elsevier}
}

@article{akram2023fog,
  title={Fog-based low latency and lightweight authentication protocol for vehicular communication},
  author={Akram, Muhammad Arslan and Mian, Adnan Noor and Kumari, Saru},
  journal={Peer-to-Peer Networking and Applications},
  volume={16},
  number={2},
  pages={629--643},
  year={2023},
  publisher={Springer}
}

@article{xiao2024calra,
  title={CALRA: practical conditional anonymous and leakage-resilient authentication scheme for vehicular crowdsensing communication},
  author={Xiao, Jianru and Ren, Yilong and Du, Jiewei and Zhao, Yanan and Kumari, Saru and Alenazi, Mohammed JF and Yu, Haiyang},
  journal={IEEE Transactions on Intelligent Transportation Systems},
  year={2024},
  publisher={IEEE}
}

@article{alladi2021deepadv,
  title={DeepADV: A deep neural network framework for anomaly detection in VANETs},
  author={Alladi, Tejasvi and Gera, Bhavya and Agrawal, Ayush and Chamola, Vinay and Yu, Fei Richard},
  journal={IEEE Transactions on Vehicular Technology},
  volume={70},
  number={11},
  pages={12013--12023},
  year={2021},
  publisher={IEEE}
}

@article{kim2022vehicular,
  title={Vehicular Multilevel Data Arrangement-Based Intrusion Detection System for In-Vehicle CAN},
  author={Kim, Wansoo and Lee, Jungho and Lee, Yousik and Kim, Yoenjin and Chung, Jingyun and Woo, Samuel},
  journal={Security and Communication Networks},
  volume={2022},
  number={1},
  pages={4322148},
  year={2022},
  publisher={Wiley Online Library}
}

@article{lee2023malicious,
  title={Malicious traffic compression and classification technique for secure Internet of Things},
  author={Lee, Yu-Rim and Park, Na-Eun and Kim, Seo-Yi and Lee, Il-Gu},
  journal={Computers, Materials and Continua},
  volume={76},
  number={3},
  pages={3465--3482},
  year={2023},
  publisher={Elsevier}
}

@article{sommer2019veins,
  title={Veins: The open source vehicular network simulation framework},
  author={Sommer, Christoph and Eckhoff, David and Brummer, Alexander and Buse, Dominik S and Hagenauer, Florian and Joerer, Stefan and Segata, Michele},
  journal={Recent advances in network simulation: the OMNeT++ environment and its ecosystem},
  pages={215--252},
  year={2019},
  publisher={Springer}
}

@inproceedings{hsu2021deep,
  title={A deep learning-based integrated algorithm for misbehavior detection system in VANETs},
  author={Hsu, Hsiao-Yuan and Cheng, Nai-Hsin and Tsai, Chun-Wei},
  booktitle={Proceedings of the 2021 ACM International Conference on Intelligent Computing and its Emerging Applications},
  pages={53--58},
  year={2021}
}

@article{slama2023one,
  title={One versus all binary tree method to classify misbehaviors in imbalanced VeReMi dataset},
  author={Slama, Omessaad and Tarhouni, Mounira and Zidi, Salah and Alaya, Bechir},
  journal={IEEE Access},
  volume={11},
  pages={135944--135958},
  year={2023},
  publisher={IEEE}
}

@article{kushardianto2024vehicular,
  title={Vehicular network anomaly detection based on 2-step deep learning framework},
  author={Kushardianto, Nur Cahyono and Ribouh, Soheyb and El Hillali, Yassin and Tatkeu, Charles},
  journal={Vehicular Communications},
  volume={49},
  pages={100802},
  year={2024},
  publisher={Elsevier}
}

@article{moulahi2023privacy,
  title={Privacy-preserving federated learning cyber-threat detection for intelligent transport systems with blockchain-based security},
  author={Moulahi, Tarek and Jabbar, Rateb and Alabdulatif, Abdulatif and Abbas, Sidra and El Khediri, Salim and Zidi, Salah and Rizwan, Muhammad},
  journal={Expert Systems},
  volume={40},
  number={5},
  pages={e13103},
  year={2023},
  publisher={Wiley Online Library}
}

@article{nagel2021white,
  title={A white paper on neural network quantization},
  author={Nagel, Markus and Fournarakis, Marios and Amjad, Rana Ali and Bondarenko, Yelysei and Van Baalen, Mart and Blankevoort, Tijmen},
  journal={arXiv preprint arXiv:2106.08295},
  year={2021}
}

@article{sajid2024enhancing,
  title={Enhancing intrusion detection: a hybrid machine and deep learning approach},
  author={Sajid, Muhammad and Malik, Kaleem Razzaq and Almogren, Ahmad and Malik, Tauqeer Safdar and Khan, Ali Haider and Tanveer, Jawad and Rehman, Ateeq Ur},
  journal={Journal of Cloud Computing},
  volume={13},
  number={1},
  pages={123},
  year={2024},
  publisher={Springer}
}

@article{sreelekshmi2025deep,
  title={A deep architecture for in-vehicle intrusion detection using controller area network-graph relied feature images},
  author={Sreelekshmi, MS and Aji, S},
  journal={Computers and Electrical Engineering},
  volume={127},
  pages={110584},
  year={2025},
  publisher={Elsevier}
}

@article{althunayyan2024robust,
  title={A robust multi-stage intrusion detection system for in-vehicle network security using hierarchical federated learning},
  author={Althunayyan, Muzun and Javed, Amir and Rana, Omer},
  journal={Vehicular Communications},
  volume={49},
  pages={100837},
  year={2024},
  publisher={Elsevier}
}

@article{tutsoy2024novel,
  title={A novel deep machine learning algorithm with dimensionality and size reduction approaches for feature elimination: thyroid cancer diagnoses with randomly missing data},
  author={Tutsoy, Onder and Sumbul, Hilmi Erdem},
  journal={Briefings in Bioinformatics},
  volume={25},
  number={4},
  pages={bbae344},
  year={2024},
  publisher={Oxford University Press}
}

\end{document}